\documentclass[12pt]{iopart}
\usepackage{hyperref}
\usepackage{graphicx}
\usepackage{bm}
\usepackage{color}

\begin{document}
\title[]{U-Net 3+ for Anomalous Diffusion Analysis enhanced with Mixture Estimates (U-AnD-ME) in particle-tracking data}

\author{Solomon Asghar$^{1,2}$, Ran Ni$^2$, and Giorgio Volpe$^1$}
\address{$^1$ Department of Chemistry, University College London, 20 Gordon Street, WC1H 0AJ London, United Kingdom}
\address{$^2$ School of Chemistry, Chemical Engineering and Biotechnology, Nanyang Technological University, Singapore,
639798}
\ead{g.volpe@ucl.ac.uk}

\begin{abstract}
Biophysical processes within living systems rely on encounters and interactions between molecules in complex environments such as cells. They are often described by anomalous diffusion transport.
Recent advances in single-molecule microscopy and particle-tracking techniques have yielded an abundance of data in the form of videos and trajectories that contain critical information about these biologically significant processes.
However, standard approaches for characterizing anomalous diffusion from these measurements often struggle in cases of practical interest, e.g. due to short, noisy trajectories. 
Fully exploiting this data therefore requires the development of advanced analysis methods -- a core goal at the heart of the recent international Anomalous Diffusion Challenges. 
Here, we introduce a novel machine-learning framework, U-net 3+ for Anomalous Diffusion analysis enhanced with Mixture Estimates (U-AnD-ME), that applies a U-Net 3+ based neural network alongside Gaussian mixture models to enable highly accurate characterisation of single-particle tracking data. 
In the 2024 Anomalous Diffusion Challenge, U-AnD-ME outperformed all other participating methods for the analysis of two-dimensional anomalous diffusion trajectories at both single-trajectory and ensemble levels. Using a large dataset inspired by the Challenge, we further characterize the performance of U-AnD-ME in segmenting trajectories and inferring anomalous diffusion properties.
\end{abstract}

\section{Introduction}
Due to its ubiquity across a broad range of fields spanning the natural sciences and beyond, diffusion has been widely studied since its first observations by Robert Brown \cite{AnomDiffRev_FO}.
According to Einstein's relation, the mean squared displacement (MSD) of Brownian motion grows linear with time $t$: ${\rm MSD} (t)  \sim Dt$, where $D$ is the diffusion coefficient \cite{Einstein1905}. Many natural and human processes show deviations from Brownian motion known as \textit{anomalous} diffusion \cite{klafter2005anomalous,AnDi_2020}, which exhibit non-linear relationships between MSD and time: ${\rm MSD} (t) \sim K t^{\alpha}$, where $K$ is the generalised diffusion coefficient and $\alpha \neq 1$ is the anomalous diffusion exponent \cite{metzler2014anomalous}.
A process is subdiffusive when $\alpha<1$, and superdiffusive when $\alpha>1$ \cite{metzler2014anomalous}.
Subdiffusion, which can occur due to crowding or interactions with boundaries, has been repeatedly observed in living cells including within cytoplasms \cite{AnomDiffCyto}, nuclei \cite{AnomDiffNuc}, and cell membranes \cite{AnomDiffMemb}.
Superdiffusion appears in active and directed systems \cite{hofling2013anomalous,ActiveAnom_B,volpe2017topography}, such as molecular motors on DNA \cite{lomholt2005optimal}. Various approaches have been recently put forward for the characterization of these processes \cite{krapf2019spectral,sposini2022towards}, also with machine-learning-based methods \cite{granik2019single,kowalek2019classification,MLforAM_Model2,munoz2020single,CONDOR,RANDI,WADNet,seckler2022bayesian,pineda2023geometric,seckler2023machine}.

Within the life sciences, advances in live-cell single-molecule imaging and particle-tracking techniques offer new insights into crucial cellular processes \cite{manzo2015review,shen2017single}.  
However, fully leveraging these technical advances  
requires further development of methods for data analysis. Often, experimental data are extracted in the form of particles' trajectories, and standard analysis methods struggle when these trajectories are, e.g., short, noisy and irregularly sampled \cite{AnDi_2020, AnDi_2024}.
Additionally, there is a need for reliable methods to identify switches between different diffusion behaviours in these trajectories, as these changes are valuable indicators of biophysical interactions within a system \cite{AnDi_2024}. Examples include variations in the generalized diffusion coefficients $K$ due to conformational changes \cite{conformational}, dimerization events \cite{dimerization} and ligand binding \cite{lingand}. Dynamics can also change due to transient immobilization \cite{immobile} and confinement effects \cite{confined}. 

With live-cell single-molecule experiments in mind, three particularly informative properties for characterizing anomalous diffusion in trajectories are $\alpha$, $K$, and the phenomenological behaviour of the diffusing particles (diffusion type), which can be classified as immobilized, confined, freely diffusing, or directed \cite{AnDi_2024}.
The international Anomalous Diffusion (AnDi) Challenges aimed to quantitatively assess the quality of existing methods for the difficult, yet important, task of identifying these properties and to spur the creation of new methods \cite{AnDi_2020, AnDi_2024}.
The last AnDi Challenge took place in 2024 and was designed specifically with biological applications in mind, focusing on two-dimensional heterogeneous diffusion in the cellular environment \cite{AnDi_2024}. 
Specifically, the Challenge aimed to evaluate methods for detecting and quantifying changes in single-particle motion, focusing on trajectory segmentation and the inference of diffusion properties at both single-trajectory and ensemble levels \cite{AnDi_2024}. The Challenge was divided into four tasks (two ensemble-level and two single-trajectory tasks) across two tracks based either on the analysis of trajectories or of videos directly.

\begin{figure*}[ht!]
\centering
\includegraphics[width=\linewidth]{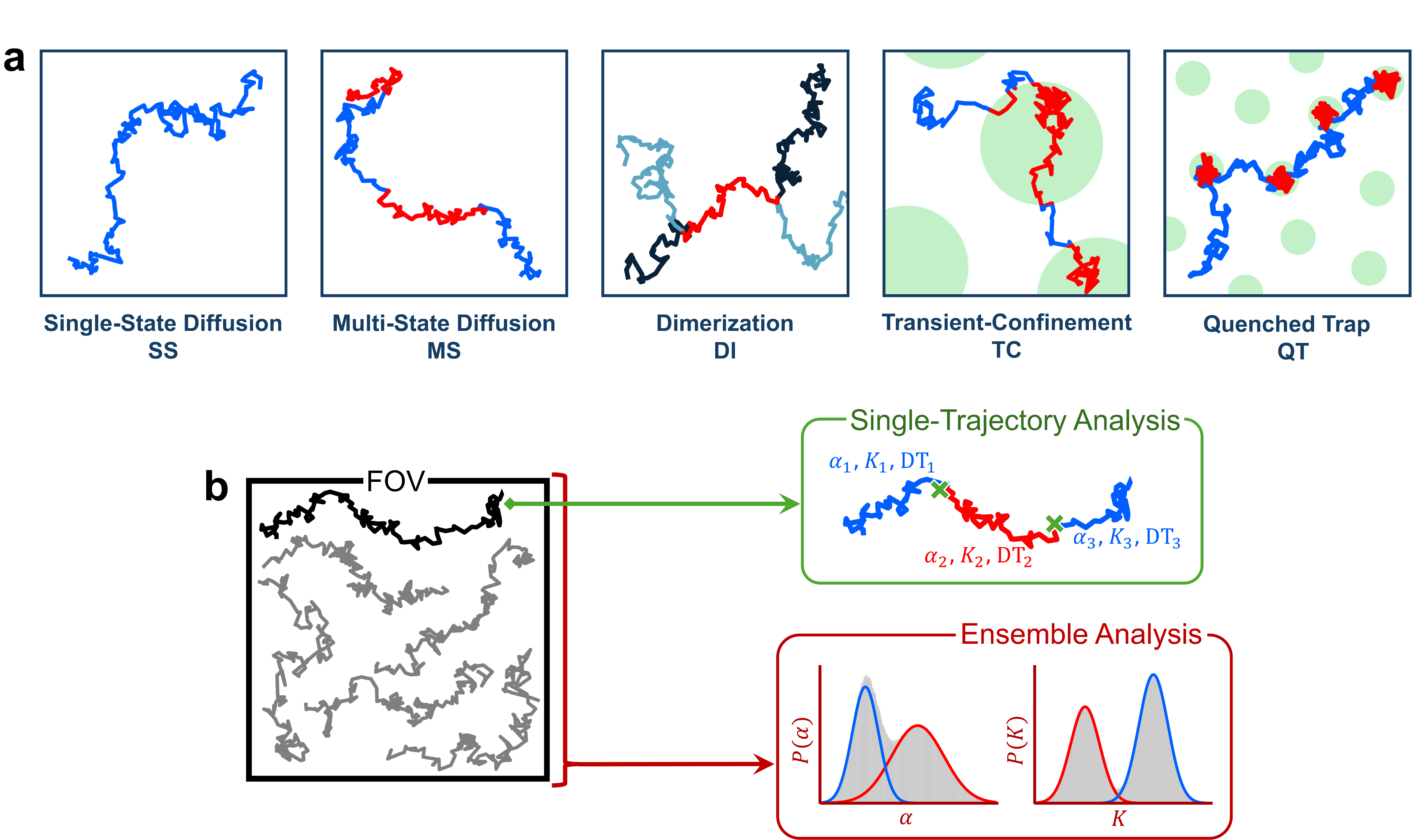}
\caption{\textbf{Overview of the 2024 AnDi Challenge.}
(\textbf{a}) The 2024 AnDi Challenge considered five physical models of diffusion \cite{AnDi_2024}
(left to right): single-state diffusion (SS) without change in properties;
multi-state diffusion (MS) spontaneously alternating between two states (red and blue); 
dimerization (DI) of two particles (light and dark blue) interacting and transiently co-diffusing (red);
the transient-confinement model (TC) with particles diffusing differently outside (blue) and inside (red) compartments (green) with osmotic boundaries; and the quenched-trap model (QT) with particles (blue) transiently immobilised (red) by traps (green).
(\textbf{b}) In the Challenge \cite{AnDi_2024}, a field of view (FOV) is composed of several trajectories (left). In the Trajectory Track, these trajectories can be analysed individually (Single-Trajectory Task) or as an ensemble (Ensemble Task).
In the Single-Trajectory Task (top right), each trajectory is analysed by detecting change points (green crosses), and, for each segment they demarcate, by inferring the anomalous-diffusion exponent $\alpha$, the generalised diffusion coefficient $K$, and the diffusion type (${\rm DT}$).
In the Ensemble Task (bottom right), analysis of an ensemble of trajectories returns the distributions for $\alpha$ and $K$, $P(\alpha)$ and $P(K)$. 
}
\label{fig:Models_and_Tasks}
\end{figure*}

Here we introduce our novel machine-learning framework for the highly accurate characterisation of anomalous diffusion properties in single-particle trajectories. We developed our framework, called U-net 3+ for Anomalous Diffusion analysis enhanced with Mixture Estimates (U-AnD-ME), to compete in the 2024 AnDi Challenge. U-AnD-ME obtained 1\textsuperscript{st} place for both tasks in the Challenge's Trajectory Track. After briefly introducing the anomalous diffusion models and metrics used for training and performance evaluation, we describe U-AnD-ME's architecture (Section \ref{sec:Method}).
Next, we benchmark our framework's performance in terms of change-point detection and inference of anomalous-diffusion properties (Section \ref{sec:Results}). 
Finally, we conclude, discussing possible future improvements and applications for U-AnD-ME (Section \ref{sec:Discussion}).

\section{Methods}\label{sec:Method}
\subsection{Anomalous Diffusion Data}\label{sec:Methods_AnDiData}

To benchmark our method against data of a known ground-truth, we simulated two-dimensional fractional Brownian motion trajectories \cite{mandelbrot1968fractional}, similar to those of the 2024 AnDi Challenge \cite{AnDi_2024}, with the \texttt{andi-datasets} Python package \cite{andiPython}. Simulations used generalized units (i.e., pixels and frames). 
The Challenge considered five different physical models of particles' motion and interaction with the environment (Fig. \ref{fig:Models_and_Tasks}a): single-state diffusion (SS) - particles have a single diffusion state \cite{singlestate}; multi-state diffusion (MS) - particles spontaneously switch between two or more diffusion states with different $K$ and/or $\alpha$ \cite{conformational,da2020heterogeneity,achimovich2023dimerization}; dimerization (DI) - particles diffuse according to a two-state model, with switching induced by random encounters with other particles \cite{lingand,valley2015enhanced,dimerization}; transient confinement (TC) - particles diffuse according to a space-dependant two-state model, being in one state when outside confined regions and the other while inside them \cite{confined,weigel2013quantifying}; and quenched trap (QT) - particles diffuse according to a space-dependant two-state model, switching between motion and immobilization by traps \cite{immobile,rossier2012integrins}. Trajectories are at most 200 frames with the minimum segment length being 3 (minimum number of time steps before a change of state or end of trajectory).

Our dataset uses a balanced composition of the same nine numerical experiments of the 2024 AnDi Challenge \cite{AnDi_2024}, where the values of the diffusion properties (Table \ref{tab:Exps}) were selected to assess the participating methods while representing biologically relevant scenarios \cite{AnDi_2024}.
Experiment 1 mimics the multi-state diffusion found in membrane proteins, with simulation parameters reproducing the three fastest states reported for the diffusion of the $\alpha$2A-adrenergic receptor \cite{Exp1_source}.
Experiment 2 reproduces changes in diffusion due to protein dimerization, as has been reported for the epidermal growth factor receptor ErbB-1 \cite{lingand}.
Experiments 3, 4, and 5 were designed to evaluate the methods' ability to detect changes from a free diffusion state to subdiffusion caused by traps, confinement regions, and dimerization, respectively.
Experiments 6 and 7 model dimerization and multi-state diffusion respectively, with both experiments using the same diffusion parameters.
Experiment 8 serves as negative control and contains only single-state diffusion trajectories with incredibly broad distributions of $\alpha$ and $K$, allowing us to test U-AnD-ME performance when diffusion properties vary significantly from its initial training distribution.
Experiment 9 is a quenched trap simulation with very short trapping times and superdiffusion in the free state. In the Challenge, for a given diffusion state, the values of the anomalous-diffusion exponent $\alpha$ and the generalised diffusion coefficient $K$ were randomly drawn from state-specific Gaussian distributions with bounds $\alpha \in (0,2)$ and $K \in [10^{-12},10^6]\ {\rm pixel}^2/{\rm frame}$, parametrized by their means ($\mu_\alpha$ and $\mu_K$) and standard deviations ($\sigma_\alpha$ and $\sigma_K$) (Table \ref{tab:Exps}). 

As in the Challenge, the structure of our dataset mirrors that of typical experimental data \cite{AnDi_2024}. Each simulated experiment is composed of three hundred fields of view (FOVs), ten times more than in the Challenge dataset. Each FOV represents a $128\times128 \ {\rm pixel}^2$ region where trajectory recording takes place, and encompasses approximately eighty trajectories on average. 
To better represent measurements from real tracking experiments, the trajectories are corrupted using Gaussian noise with zero mean and a standard deviation $\sigma=0.12\ {\rm pixels}$. 
Particles within the same FOV can interact with one another and/or with the FOV's environment. 

\begin{table}[h!]
\begin{center}
\begin{tabular}{l l r r r r r r}
    \hline
    Exp. & Model & $\mu_\alpha$ & $\sigma_\alpha$  & $\mu_K$ & $\sigma_K$  & Weight \\
    \hline
    1          & MS    & 1.00         & 0.0001        & 0.15    & 0.01     & 0.30   \\
               &       & 1.00         & 0.01          & 0.33    & 0.001    & 0.49   \\
               &       & 1.00         & 0.01          & 0.95    & 0.01     & 0.21   \\
    \hline
    2          & DI    & 1.00         & 0.1           & 0.28    & 0.001    & 0.76   \\
               &       & 1.10         & 0.01          & 0.0035  & 0.0001   & 0.24   \\
    \hline
    3          & QT    & 0.00         & 0.0           & 0.0     & 0.0      & 0.15   \\
               &       & 1.00         & 0.005         & 1.0     & 0.1      & 0.85   \\
    \hline
    4          & TC    & 0.20         & 0.001         & 0.01    & 0.001    & 0.56   \\
               &       & 1.00         & 0.005         & 1.0     & 0.1      & 0.44   \\
    \hline
    5          & DI    & 0.20         & 0.001         & 0.01    & 0.001    & 0.31   \\
               &       &  1.00        & 0.005         & 1.0     & 0.1      & 0.69   \\
    \hline
    6          & DI    & 0.70         & 0.1           & 0.1     & 0.1      & 0.86   \\
               &       & 1.20         & 0.001         & 1.0     & 0.01     & 0.14   \\
    \hline
    7          & MS    & 0.70         & 0.1           & 0.1     & 0.1      & 0.74   \\
               &       & 1.20         & 0.01          & 1.0     & 0.01     & 0.26   \\
    \hline
    8          & SS    & 1.00         & 10            & 1.0     & 100      & 1.00   \\
    \hline
    9          & QT    & 0.00         & 0.0           & 0.0     & 0.0      & 0.58   \\
               &       & 1.99         & 0.01          & 1.0     & 0.01     & 0.42   \\
    \hline
\end{tabular}
\end{center}
\caption{\textbf{Simulated experimental properties.} Columns show the diffusion model of each numerical experiment, along with the diffusion properties (mean $\mu$ and standard deviation $\sigma$ of the anomalous-diffusion exponent $\alpha$ and the generalized diffusion coefficient $K$) and weights of each diffusion state. 
}
\label{tab:Exps}
\end{table}

In the Trajectory Track of the Challenge, experiments could be analysed in two distinct ways \cite{AnDi_2024}, based on predictions at either single-trajectory (Single-Trajectory Task) or ensemble level (Ensemble Task) (Fig. \ref{fig:Models_and_Tasks}b). Our framework allows for both types of predictions. Single-trajectory predictions involve the detection of all the change points within a trajectory and, for each segment these change points demarcate, the inference of $K$, $\alpha$ and diffusion type (DT) -- an identifier of what kind of constraint is imposed by the environment (immobile = 0, confined = 1, free = 2, directed = 3). Ensemble predictions describe each experiment collectively, capturing the distributions of $\alpha$ and $K$ across all of its trajectories.

\subsection{Evaluation Metrics}\label{sec:Methods_evaluation}
Single-trajectory predictions used two different metrics to assess the detection of change points \cite{AnDi_2024}: The Jaccard similarity coefficient (JSC\textsubscript{CP}, Eqn. \ref{JSC}) and the root mean squared error (RMSE\textsubscript{CP}, Eqn. \ref{RMSE}).
Given a ground-truth change point at location $t_{{\rm GT},i}$ (with $i$ an integer) and a detection at location $ t_{{\rm P},j}$ (with $j$ an integer), a gated absolute distance is defined as
$d_{i,j} = {\rm min}(|t_{{\rm GT},i} - t_{{\rm P},j}|, \varepsilon_{\rm CP})$, where $\varepsilon_{\rm CP}=10$ is a fixed penalty for change points more than $\varepsilon_{\rm CP}$ apart.
The number of detected change points may not always match the number of true ones. 
In these cases, change points were  assigned as a rectangular assignment problem using the Hungarian algorithm \cite{Hungarian} by minimising the sum of distances between paired change points, $d_{\rm CP} = \mathop{{\rm min}}_{{\rm paired \, CP}}\big(\sum d_{i,j} \big)$.
After this assignment, we calculated the number of true positive (TP), false positive (FP) and false negative (FN) detections. 
A detection was considered a TP if it was within $\varepsilon_{\rm CP}$ of its paired ground-truth value.
Predictions not associated with any ground-truth values or more than $\varepsilon_{\rm CP}$ away from their assigned value were considered FP.
Ground-truth change points with no assigned detection within $\varepsilon_{\rm CP}$ were considered FN.
The overall number of TP, FP and FN was used to calculate the Jaccard similarity coefficient for the change-point detections over an experiment:
\begin{equation}
{\rm JSC}_{\rm CP} = \frac{{\rm TP}}{{\rm TP} + {\rm FN} + {\rm FP}}
\label{JSC}
\end{equation}
${\rm JSC}_{\rm CP}$ takes values between 0 and 1, with 1 being a perfect score.
The root mean squared error of the TP detections was also calculated:
\begin{equation}
{\rm RMSE}_{\rm CP} = \sqrt{\frac{1}{N} \sum_{\rm TP} \big(t_{{\rm GT},i} - t_{{\rm P},j} \big)^2}
\label{RMSE}
\end{equation}
where $N$ is the number of TPs.
Lower ${\rm RMSE}_{\rm CP}$ values indicate detection with lower localization error.
Together, the two metrics quantified the quality of change-point predictions in terms of both accuracy and resolution \cite{AnDi_2024}.

After identifying change points, inference of $\alpha$, $K$ and diffusion type (DT) can be made for each segment they delineate. 
For the $N$ paired segments, inference of the anomalous-diffusion exponent $\alpha$  was evaluated via a mean absolute error (MAE):
\begin{equation}
{\rm MAE}_\alpha = \frac{1}{N} \sum_{\rm seg} | \alpha_{{\rm GT},i} - \alpha_{{\rm P},j}|
\label{MAE}
\end{equation}
where $\alpha_{{\rm GT},i}$ and $\alpha_{{\rm P},j}$ are the ground-truth and predicted values of $\alpha$, respectively. Evaluation of the generalised diffusion coefficient $K$ used the mean squared logarithmic error (MSLE):
\begin{equation}
{\rm MSLE}_{K} = \frac{1}{N} \sum_{\rm seg}\big[\log(K_{{\rm GT},i} + 1) -  \log(K_{{\rm P},j} + 1) \big]^2
\label{MSLE}
\end{equation}
where $K_{{\rm GT},i}$ and $K_{{\rm P},j}$ are the ground-truth and predicted values of $K$, respectively.
\noindent Lower values of ${\rm MAE}_\alpha$ and ${\rm MSLE}_{K}$ indicate better predictions.
The diffusion type (DT) was evaluated using the ${\rm F}_1$-score:
\begin{equation}
 {\rm F}_1 = \frac{2{\rm TP}_{\rm c}}{2{\rm TP}_{\rm c} + {\rm FP}_{\rm c} + {\rm FN}_{\rm c}}
\label{F1}
\end{equation}
\noindent where ${\rm TP}_{\rm c}$, ${\rm FP}_{\rm c}$ and ${\rm FN}_{\rm c}$ are the true positives, false positives, and false negatives with respect to segment classification.
Due to the presence of class imbalance, this metric is calculated as a micro-average which aggregates the contributions of all classes \cite{AnDi_2024}.
${\rm F}_1$-score takes values between 0 and 1, with 1 being the best possible score.

Finally, ensemble predictions were evaluated using the estimated mean, standard deviation, and relative weight of each state's $\alpha$ and $K$ to define the multimodal distributions $P(\alpha)$ and $P(K)$ (Fig. \ref{fig:Models_and_Tasks}b). The similarity of these distributions to the ground-truth distributions $Q(\alpha)$ and $Q(K)$ (Table \ref{tab:Exps}) was assessed using the first Wasserstein distance \cite{AnDi_2024}:
\begin{equation}
	W_1(P,Q) = \int_{{\rm supp}(Q)} |{\rm CDF}_P(x) - {\rm CDF}_Q(x)|dx
\label{Wasserstein}
\end{equation}
\noindent where ${\rm CDF}$ refers to a distribution's cumulative distribution function and ${\rm supp}(Q)$ is the support, i.e. $\alpha \in (0,2)$ or $K \in [10^{-12}, 10^6]\ {\rm pixel}^2/{\rm frame}$.
$W_1(P,Q)$ approaches 0 as the accuracy of the predictions improves.
Henceforth, we refer to the first Wasserstein distance of $\alpha$ as $W_{\alpha}$ and of $K$ as $W_{K}$.

\subsection{U-AnD-ME Framework}\label{sec:Methods_UAndME}

\begin{figure*}[h]
\centering
\includegraphics[width=\linewidth]{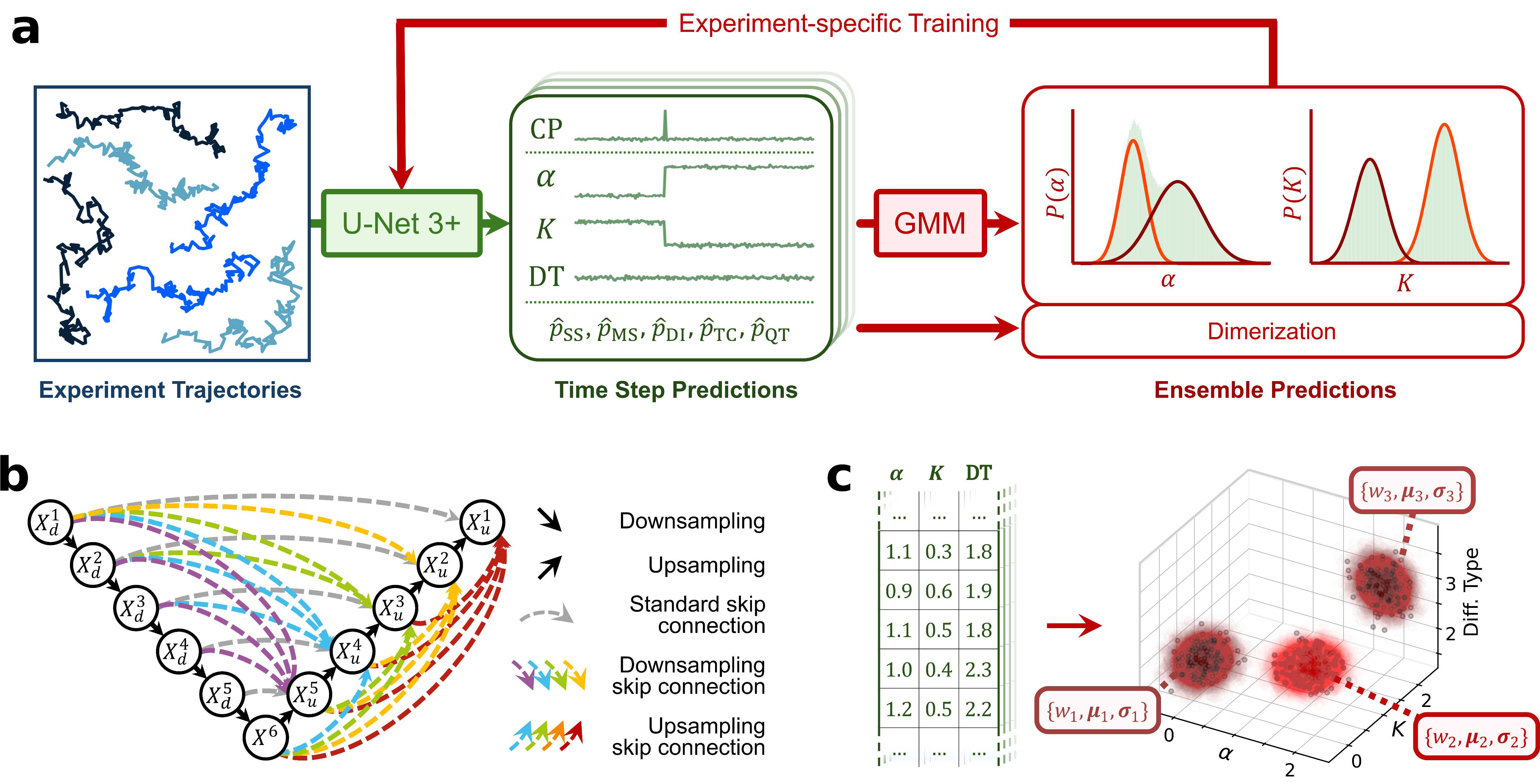}
\caption{\textbf{U-AnD-ME workflow.}
(\textbf{a}) A network inspired by U-Net 3+ processes each trajectory of an experiment. For each time step, it predicts the probability of it being a change point (CP), $\alpha$, $K$, diffusion type (DT), and the likelihood of belonging to each of the five diffusion models ($\hat{p}_{\rm SS}$, $\hat{p}_{\rm MS}$, $\hat{p}_{\rm DI}$, $\hat{p}_{\rm TC}$, $\hat{p}_{\rm QT}$). These time-step predictions are processed to produce trajectory-level predictions (not pictured).
All diffusion model predictions in an experiment are averaged to predict its most likely model, here dimerization.
Additionally, all the $\alpha$, $K$ and DT predictions in an experiment are used to create a Gaussian Mixture Model (GMM) and estimate the probability distributions $P(\alpha)$ and $P(K)$.
These ensemble predictions inform the training a second U-Net 3+ inspired network, making it experiment-specific and thereby more accurate.
(\textbf{b}) Schematic of a U-Net 3+ architecture. Each $X_d$ ($X_u$) is a node in the downsampling (upsampling) branch. $X^6$ is the bridge between the two branches. Solid downwards (upwards) black arrows represent downsampling (upsampling). Dashed gray arrows show standard skip connections. Dashed  downwards (upwards) coloured arrows represent downsampling (upsampling) skip connections.
The colour of these arrows indicates their output shape based on their end-point node: for example, each green arrow reshapes its input to match the size of $X_u^3$.
(\textbf{c}) The predictions of $\alpha$, $K$ and diffusion type (DT) for each time step of every trajectory in an experiment inform the creation of 
a three dimensional GMM (using diagonal covariance only). The parameters of each component of the GMM (weight $w$, mean $\mu$ and standard deviations $\sigma$) represent a different diffusion state, and, collectively, capture that experiment's ensemble properties. In the example, three components are identified. 
}
\label{fig:UAnDME}
\end{figure*}

U-AnD-ME (Fig. \ref{fig:UAnDME}a) processing begins by using a neural network based on U-Net 3+ \cite{UNet3+} (Fig. \ref{fig:UAnDME}b) to make predictions for each time step in a trajectory (Section \ref{sec:Architecture}).
Before being passed to this network, trajectories must be preprocessed (Section \ref{sec:pre-proprocessing}).
The initial neural network is trained to handle a broad range of experimental conditions, and so we refer to it as the \textit{generalist} network (Section \ref{sec:GeneralistTraining}).
For each time step, this network predicts the likelihood of it being a change point, estimates its $\alpha$, $K$, and diffusion type (DT), and its likelihood of belonging to each of the five possible diffusion models.
These time-step predictions are used to segment each trajectory and label the properties of each segment (Section \ref{sec:single_traj_preds}), solving the single trajectory task of the 2024 AnDi Challenge.
Additionally, to make ensemble predictions (Section \ref{sec:EnsemblePreds}), the combined time-step predictions from all the trajectories in an experiment can be used to infer the most likely diffusion model behind that experiment and for the creation of a Gaussian Mixture Model (GMM) capturing the experiment's $\alpha$- and $K$-distributions (Fig. \ref{fig:UAnDME}c).
This ensemble information can be leveraged to further refine U-AnD-ME predictions (Section \ref{sec:exp_specific_train}).
Ensemble predictions are used to generate trajectories representative of each experiment, enabling training of a new, more accurate \textit{experiment-specific} network, also based on U-Net 3+ \cite{UNet3+} (Fig. \ref{fig:UAnDME}b). 
We implemented our framework in Python 3 using TensorFlow 2 and NumPy. All codes pertaining to U-AnD-ME are freely available under an MIT licence \cite{UAnDMEZenodo}. We chose all parameters of our architecture by optimizing the metrics of the 2024 AnDi Challenge. All computations used the same node on the Gekko cluster of Nanyang Technology University with an Intel Skylake Xeon Gold 6150 processor and Nvidia V100 GPU for network training.

\subsubsection{Architecture.}\label{sec:Architecture}

U-Net 3+ (Fig. \ref{fig:UAnDME}b) is the inspiration behind U-AnD-ME's central neural network (for both generalist and experiment-specific cases). It is a convolutional architecture, originally developed for biomedical image segmentation, consisting of an encoding/downsampling branch (Fig. \ref{fig:UAnDME}b, left) followed by a decoding/upsampling branch (Fig. \ref{fig:UAnDME}b, right), with complex skip connections interlinking the two \cite{UNet3+}. 

Although U-Net architectures were originally formulated for image-segmentation tasks, their hierarchical approach to feature extraction has since been applied to time series analysis \cite{UTime, USleep}, including to the analysis of anomalous diffusion data \cite{UAnDi}.
To adapt the architecture to time series, we replaced the two-dimensional convolutions of the original architecture with one-dimensional convolutions.  
The downsampling branch compresses the input and extracts coarse-grained semantic information, while the upsampling branch combines this semantic information with fine-grained details from the skip connections, enabling context-aware processing of the input. Both branches are roughly symmetric leading to its eponymous `U' shape. As the shape of the data changes at each step, the nodes $X$ of both branches are sometimes referred to as \emph{scales} \cite{UNet3+,UNet++}.
In Fig. \ref{fig:UAnDME}b, the node $X_d^1$ ($X_u^1$) is the largest scale of the downsampling (upsampling) branch, while $X_d^5$ ($X_u^5$) is the smallest scale. $X^6$ is the bridge between the two branches. 
We implement six scales in total, as this is the maximum possible with our input length. In fact, our network accepts 224 $\times$ 2 (time steps $\times$ dimensions) matrices as input, with any shorter trajectories padded to this length as it allows up to $N_{\rm d}=5$ downsampling operations, thus enabling deeper and more expressive networks.

The downsampling branch follows a standard architecture for convolution networks, being composed of scales implementing repeated (valid) one-dimensional convolutions with a kernel size of 3 and stride of 1, each followed by a rectified linear unit (ReLU) activation function and a one-dimensional max-pooling operation. The latter operates along the time axis with a pooling size of 2 and a stride of 2, thus halving the data dimensionality at each downsampling step. After each of these downsampling steps the number of feature channels (i.e., vectors abstractly encoding extracted information) increases. We set the number of channels for $X_{d}^1$, $X_{d}^2$, $X_{d}^3$, $X_{d}^4$, $X_{d}^5$ and $X^6$ to 16, 32, 64, 64, 128 and 128, respectively (Fig. \ref{fig:UAnDME}b).
Every scale in the upsampling branch consists of a 1D transposed convolution along the time axis with a kernel size of 2 and a stride of 2; this operation doubles the data dimensionality, inverting the shape changes caused by the downsampling operations. 
Each transposed convolution in our upsampling branch uses 512 channels.
Skip connections allow the output of each up-convolution to be combined with features from the downsampling branch.
Every node of the upsampling branch incorporates information from its same-scale counterpart from the downsampling branch and, additionally, from all larger-scale downsampling nodes, and from all smaller-scale upsampling nodes including the bridge (Fig. \ref{fig:UAnDME}b).
This means that the skip connections must also include downsampling and upsampling operations as appropriate, reshaping their input to match the shape of the upsampling branch they connect to.
Incorporating skip connections that combine scales in this way enhances the integration of coarse-grained semantic information with finely detailed information, allowing the network to better understand the context of the input. 

Finally, a $1\times1$ convolution operates on the output $X_u^1$ to ensure that the number of features matches the desired number of outputs: in our case, this convolution reduces the number of channels from $512$ to 9, making the shape of the final output $224\times9$ and encoding the nine predicted feature channels for each of the 224 time steps.
The first channel undergoes a sigmoid activation and represents the presence of change points.
The next three have no activation function and represent $K$, $\alpha$ and diffusion type (DT).
The remaining five channels  undergo a five-way softmax activation, and represent the probability of a time step to belong to each of the five possible phenomenological diffusion models.
Experiment-specific networks can also be created once experimental ensemble properties have been predicted once (Fig. \ref{fig:UAnDME}a).
As the diffusion model is fixed for each experiment, these networks do not need to make any further model prediction, so that the $1\times1$ convolution after $X_{u}^1$ reduces the number of channels to just 4 instead of 9 (presence of change point, $\alpha$, $K$ and DT).

\subsubsection{Trajectory Pre-processing.}\label{sec:pre-proprocessing}

\begin{figure*}[ht]
\centering
\includegraphics[width=\linewidth]{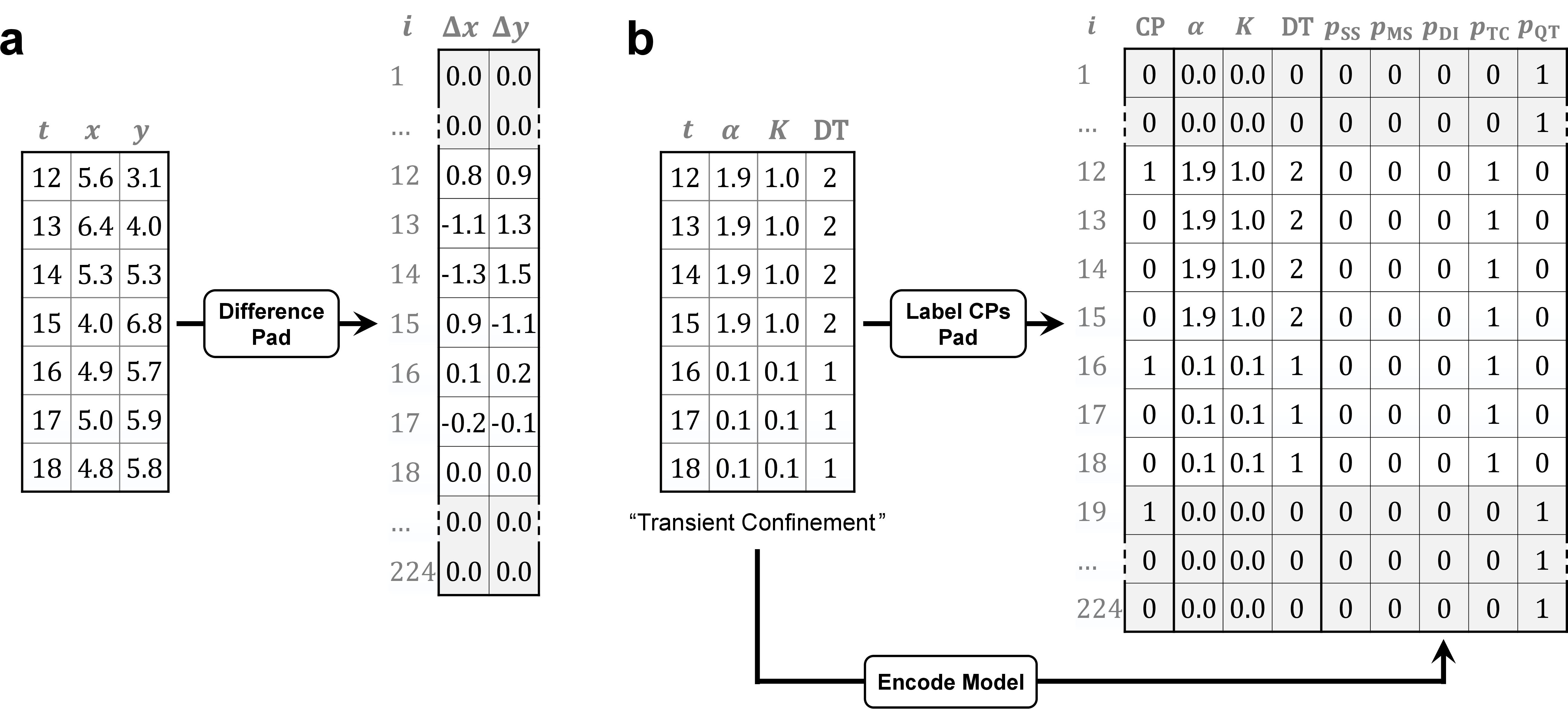}
\caption{\textbf{Trajectories pre-processing.}
(\textbf{a}) Raw trajectories (left) consist of time-step labels $t$ and positions, $x$ and $y$, for each time step. 
They are differenced in time to yield the increments $(\Delta x_t,\Delta y_t) = (x_{t+1},y_{t+1}) - (x_{t},y_{t})$.
Processed trajectories (right) do not contain explicit time labels $t$, but their index $i$ implicitly captures this information, i.e. $(\Delta x_i,\Delta y_i) \equiv (\Delta x_t,\Delta y_t)$.
Missing values are padded with zeros (gray) to a fixed length of 224 time steps.
In the example, the original trajectory spans $t=12$ to $t=18$ (time when it was within the FOV). The increments $(\Delta x_t,\Delta y_t)$ are therefore defined for $i=12$ to $i=17$, while zeros are used to fill the otherwise undefined values, from $i=1$ to $i=11$ and from $i=18$ to $i=224$. 
(\textbf{b}) Corresponding raw labels (left) consist of the same time-step labels $t$, with the respective values of $\alpha$, $K$, and DT. Additionally, a label describes the diffusion model of the trajectory, here ``Transient Confinement".
For both generalist and experiment-specific networks, raw labels are processed (right) by adding explicit change point labels CP (set to one for the start of a new segment and zero otherwise) for each time step. As for trajectories, the indices $i$ of the new matrix captures time information. Moreover, for generalist networks (as in the depicted example), the diffusion model is one-hot encoded and also forms part of the label through the probabilities $p_{\rm SS}$, $p_{\rm MS}$, $p_{\rm DI}$, $p_{\rm TC}$ and $p_{\rm QT}$.
In the example, for each original time step ($i=12$ to $i=18$), the probability of belonging to the transient confinement mode, $p_{\rm TC}$, is set to one, while all other probabilities are set to zero.
CP, $\alpha$, $K$, and DT values are padded with zeros (gray). 
In generalist networks, model labels are padded with zeros too bar the probability of being in a quenched trap, $p_{\rm QT}$, which is set to one. This padding mimics an immobilizing trap at the FOV edge.}
\label{fig:Preprocessing}
\end{figure*}

Before being fed into a network, trajectories (of maximum 200 frames) were preprocessed to simplify learning and ensure appropriate tensor sizes.
Raw trajectories consist of explicit time-step labels $t$, with position values ($x$ and $y$) for each time step (Fig. \ref{fig:Preprocessing}a). They 
may not span from 1 to 200, as particles may enter a FOV after $t=1$ and leave before $t=200$.
These trajectories are first differenced in time to yield increments $(\Delta x_t,\Delta y_t) = (x_{t+1},y_{t+1}) - (x_{t},y_{t})$. Processed trajectories do not contain explicit time labels $t$, but their index $i$ implicitly captures this information, i.e. $(\Delta x_i,\Delta y_i) \equiv (\Delta x_t,\Delta y_t)$.
Missing values are padded with zeros to a fixed length of 224 time steps (Section \ref{sec:Architecture}).

Corresponding raw labels (Fig. \ref{fig:Preprocessing}b) consist of the same time steps $t$, along with values of $\alpha$, $K$, and diffusion type (DT).
Additionally, a label describes the diffusion model behind the trajectory.
For both generalist and experiment-specific networks, the labels for $\alpha$, $K$ and DT are used without any additional processing. As DT is an ordinal category, we simply treat it as a float variable. As for trajectories, the label index $i$ implicitly encodes the original time information $t$. Padding sets the missing values of $\alpha$, $K$ and DT to zero up to the fixed length of 224 time steps of the processed trajectories. Processed labels include explicit change point information (CP) too, which is set to one for the start of any segment and for the first time step after the end of the raw trajectory; it is set to zero otherwise. 
Unlike experiment-specific networks, generalist networks also require information about the diffusion model. As this information is not ordinal, we apply one-hot-encoding: each time step has five labels encoding whether it belongs to each of the five diffusion models; the label for the true model is set to 1 while all others are 0.
For padded time steps, the diffusion model is set as a quenched trap, i.e. $p_{\rm QT}=1$ while $p_{\rm SS}=p_{\rm MS}=p_{\rm DI}=p_{\rm TC}=0$.
This padding scheme essentially treats the boundary of every FOV as an immobilizing trap, preventing the padding from adding any physically unrealistic behaviour to the training data.

\subsubsection{Network Training.}\label{sec:GeneralistTraining}

We simulated fractional Brownian motion trajectories for training using the \texttt{andi-datasets} Python package \cite{andiPython}. 
For generalist networks, we simulated trajectories corresponding to all five diffusion models, with all their parameters randomly selected from a predefined range informed by the AnDi 2024 pilot dataset \cite{AnDi_2024}.  
Due to the negative control nature of Experiment 8, $K$-values can purposely span a very broad range. Training a network over such a large range would be computationally expensive and result in generally poor performance. We therefore limited the range for training the generalist networks to the values more represented in the Challenge dataset (Table \ref{tab:Exps}) \cite{AnDi_2024}. 
Trajectories were therefore simulated for each diffusion state in an experiment with values of $\alpha$ and $K$ sampled from Gaussian distributions with means, $\mu_\alpha \sim {\rm U}(0, 1.999)$ and $\mu_K \sim {\rm U}(10^{-12}, 15)$, and standard deviations, $\sigma_\alpha = 0.01\mu_\alpha$ and $\sigma_K = 0.01\mu_K$. 
SS diffusion requires only one state.
For MS, we simulated a maximum of five states, as this was comfortably larger than the maximum of three found in the pilot dataset \cite{AnDi_2024}.
MS also requires the definition of a transition matrix $M$ \cite{AnDi_2024}. At any time step, the transition probability from a state $i$ to a state $j$ is given by $M_{ij}$.
The probability of remaining in the same state $i$ is given by $M_{ii}$, which we set to a single value $M_{ii} \sim {\rm U}(0.9, 0.999)$ for all states. The values for all other transition probabilities are then $M_{ij}=(1-M_{ii})/(s -1)$, where $s$ is the number of states.
The DI model requires two states and the definition of the number of diffusing particles $N$, their radius $r$, the probability $P_{\rm b}$ that two particles bind when at a distance $d<2r$, and the probability $P_{\rm u}$ of a dimer unbinding.
We generated experiments for this model using $N \sim {\rm U}(50,150)$, $r \sim {\rm U}(0.1,2.0)$, $P_{\rm b} = 1$, and $P_{\rm u} \sim {\rm U}(0,0.1)$.
The TC model also requires two states and the definition of the number of confinement regions $N_{\rm c}$, their radius  $r_{\rm c}$, and the transmittance probability of the boundary $T$.
We used $N_{\rm c} \sim {\rm U}(10,100)$, $r_{\rm c} \sim {\rm U}(1,15)$, and $T \sim {\rm U}(0,0.5)$.
Finally, the QT model requires only the definition of one state, as the other is complete immobilization ($\mu_\alpha = \mu_K = \sigma_\alpha = \sigma_K = 0$).
It also requires the number of traps $N_{\rm t}$, their radius $r_{\rm t}$, the trap binding probability $P_{\rm t}$ when at a distance $d<r_{\rm t}$, and the unbinding probability $P_{\rm u}$.
We used $N_{\rm t} \sim {\rm U}(100,500)$, $r_{\rm t} \sim {\rm U}(0.1,2.0)$, $P_{\rm b} = 1$, and $P_{\rm u} \sim {\rm U}(0,0.1)$.

The training of both generalist and experiment-specific networks followed the same procedure, but the simulation parameters used for experiment-specific training came directly from the ensemble predictions of the generalist network for that experiment (see also Section \ref{sec:exp_specific_train}). 
As the networks' output is multimodal, we used several different loss functions in unison for training: a binary cross-entropy loss for the binary classification task of detecting change points; mean squared error losses for the regression tasks of inferring $\alpha$, $K$ and diffusion type (DT); and, only for generalist networks, categorical cross-entropy for the multi-class classification task of predicting the diffusion model.
Training proceeded via training loops that consisted of the generation of 50,000 new trajectories (40,000 for training and 10,000 for validation). Each loop proceeded until the validation loss stagnated for five consecutive epochs. Training came to a final stop when the minimum validation loss per loop stopped improving between loops. We finally selected the networks parameters from the overall validation minimum. 
Both generalist and experiment-specific networks reached their validation minimum after ca. 18 hours on our computational resources.

\subsubsection{Single-Trajectory Predictions.}\label{sec:single_traj_preds}

\begin{figure*}[ht]
\centering
\includegraphics[width=0.7\linewidth]{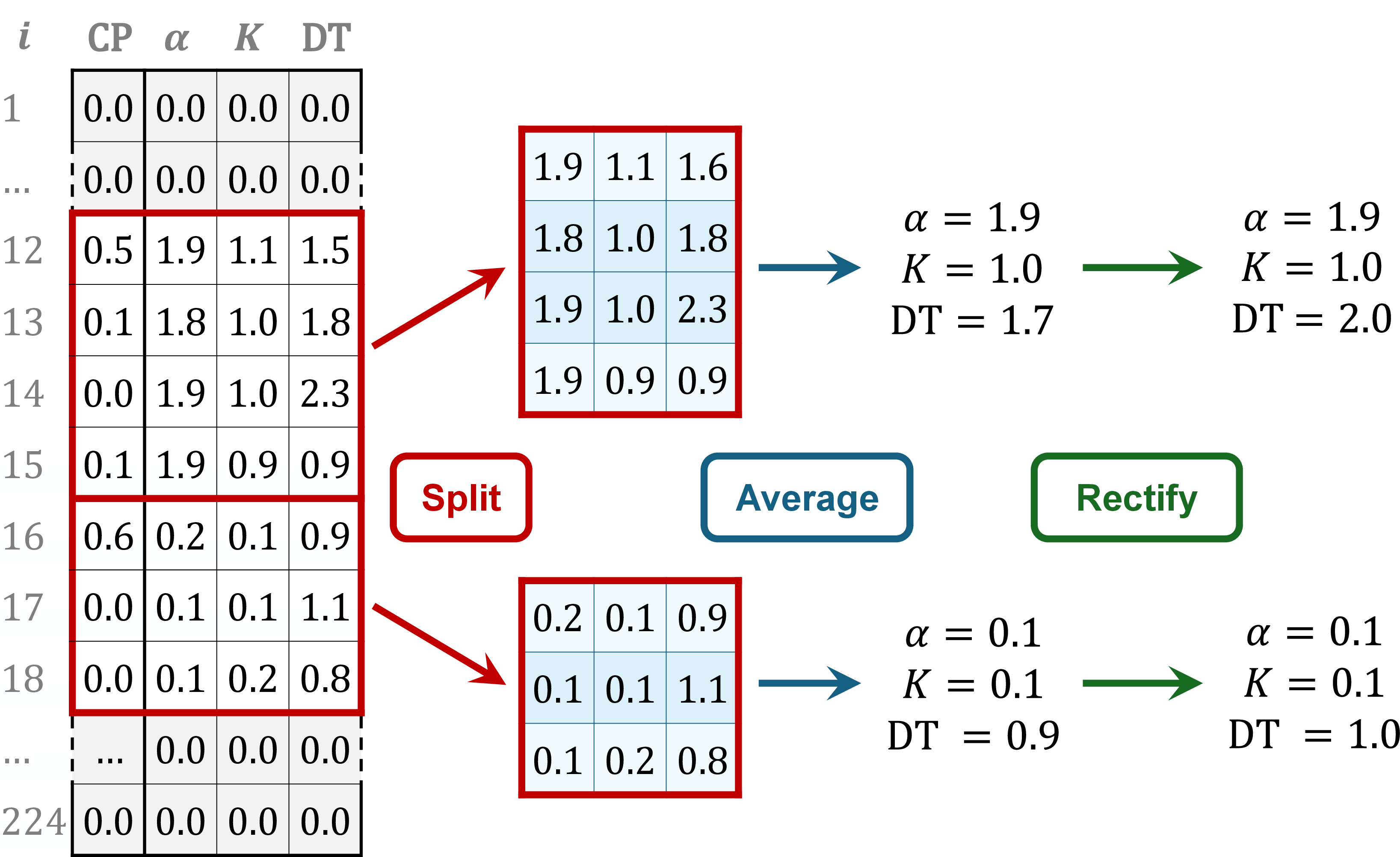}
\caption{\textbf{U-AnD-ME single-trajectory prediction procedure.}
The network output consists of predictions of change point (CP) probability, $\alpha$, $K$, and diffusion type  (DT) for each of the 224 time steps of the processed trajectories. This output includes padded values (gray). 
After padding is reverted, the output is split into segments using the change point predictions (red boxes).
Each identified segment then undergoes a weighted average (blue shades), with its central time steps assigned higher weights, which yields estimates for $\alpha$, $K$, and DT.
Finally, these values are rectified, being constrained and rounded as appropriate.
}
\label{fig:SingleTrajPred}
\end{figure*}

The single-trajectory prediction procedure was identical for generalist and experiment-specific networks (Fig. \ref{fig:SingleTrajPred}).  
For each trajectory, after removing padded time steps, change points were detected first. We considered a time step to be a change point if its change-point label was at least 0.25 and a local maximum compared to its immediate neighbours. Change points within two time steps of the start or end of the unpadded output were ignored, as the minimum possible segment length was three. The output tensor was split into segments according to these identified change points (Fig. \ref{fig:SingleTrajPred}). 
The values of $\alpha$, $K$ and diffusion type (DT) for each time step in a segment were averaged to generate a singular prediction for that segment. 
This average used a parabolic weighting, where time steps near the centre of the segment contributed more than those at its extremities as errors in the change-point localisation make them less reliable (Fig. \ref{fig:SingleTrajPred}).
For a segment spanning $i_{\rm start}$ to $ i_{\rm end}$, each time step's weighting $w_i$ is given by 
$w_i = \frac{3}{10}(2 - \tilde{i}^{2})$,
where $\tilde{i} = 2 \frac{i - i_{\rm start}}{i_{\rm end} - i_{\rm start}} - 1$ is a mapping of $i$ from $[i_{\rm start}, i_{\rm end}]$ to $[-1,1]$ and the prefactor $\frac{3}{10}=(\int^1_{-1}2-\tilde{i}^{2}\ d\tilde{i})^{-1}$ is a normalisation factor ensuring that $\sum_{i=i_{\rm start}}^{i_{\rm end}} w_i=1$. Finally, prediction of segment properties were rectified (Fig. \ref{fig:SingleTrajPred}) by constraining all values to within physically possible/realistic ranges, and ensuring they were of an appropriate data type.
Predictions for $\alpha$ and $K$ were constrained to $(0,2)$ and $[10^{-12},10^6]$, while those for the diffusion type were rounded to the nearest integer and then constrained to $[0,3]$ (Section \ref{sec:Methods_AnDiData}).

\subsubsection{Ensemble Predictions.}\label{sec:EnsemblePreds}

Ensemble predictions aim at approximating each experiment's multimodal distributions of $\alpha$ and $K$, $P_\alpha$ and $P_K$ (Table \ref{tab:Exps}, Fig. \ref{fig:UAnDME}a). 
Using standard expectation maximisation \cite{GMM_EM_algo}, 
we fitted a Gaussian Mixture Model (GMM) to the joint distribution of all values of $\alpha$, $K$ and diffusion type (DT) predicted by our generalist network for all time steps of the trajectories in an experiment  (Fig. \ref{fig:UAnDME}c).
In our case, we used this GMM to approximate the multimodal distributions of $\alpha$, $K$, and diffusion type as a sum of Gaussian components, where each component $i$ is characterized by means $\mu_{\alpha,i}$, $\mu_{K,i}$, and $\mu_{{{\rm DT},i}}$, standard deviations $\sigma_{\alpha,i}$, $\sigma_{K,i}$, and $\sigma_{{{\rm DT},i}}$, and a weight $w_{i}$. While the properties of the DT distribution are not strictly necessary, considering this property led to better separation between different diffusion states, and, thus to a better accuracy for the captured distributions $P_\alpha$ and $P_K$.
Our GMM used strictly diagonal covariance matrices, meaning the three variables ($\alpha$, $K$ and DT) are independent and have different standard deviations. In the ideal case, the number of Gaussian components will match that of diffusion states in an experiment exactly. In practice, this is rarely the case, yet we found that the overall GMM distributions over $\alpha$ and $K$ still closely approximate the experimental distributions.
We created GMMs with between one and ten components, and finally selected that with the lowest Bayesian information criterion \cite{GMM_OptTeqs}.
We set ten as the maximum number of components as this led to strong performance with good efficiency. Finally, we also used the outputs of the generalist network to predict the probability of an experiment to belong to each of the five possible diffusion models. We used this information to train our experiment-specific networks together with the probabilities $P_\alpha$ and $P_K$ defined by the GMM (Section \ref{sec:exp_specific_train}). To avoid error propagation from poor diffusion model predictions, we never used this information to decide the number of GMM components.

\subsubsection{Experiment-Specific Networks}\label{sec:exp_specific_train}
Once approximate experimental properties were available from the predictions of generalist networks, we trained and used experiment-specific networks to improve accuracy for both single-trajectory and ensemble predictions.
Training proceeded as outlined in Section \ref{sec:GeneralistTraining} but with new trajectories simulated in accordance with the predicted diffusion model using ranges of $\alpha$ and $K$ defined by the distributions predicted by the generalist GMM (Section \ref{sec:EnsemblePreds}). 

Ideally, a single model will be predicted for the experiment with high confidence.
However, often multiple models show comparable probabilities, so it is important not to overcommit to just the most probable one, as this risks training the network on the wrong model. Therefore,
we trained experiment-specific networks using trajectories from multiple models selected from the softmax output $\{\hat{p}_{\rm SS}, \hat{p}_{\rm MS}, \hat{p}_{\rm DI}, \hat{p}_{\rm TC}, \hat{p}_{\rm QT}\}$ of the generalist network encoding the probability of the experiment being SS, MS, DI, TC or QT, respectively. We sorted this set of probabilities into descending order and calculated the difference between any two consecutive probabilities in the sorted list.
Training used all models up to the first where this difference exceeded the predefined threshold value of 0.1. 
We used this value because it stroke an acceptable balance between accuracy (the generated set of diffusion models generally included the true model) and specificity (the generated set of diffusion models was small) on a trial dataset.

Simulating trajectories of an experiment for training requires the definition of $\mu_\alpha$, $\mu_K $, $\sigma_\alpha$ and $\sigma_K$ for each diffusion state in it (Section \ref{sec:GeneralistTraining}).
While the training of the generalist network uses random values for these parameters, when training experiment-specific networks, these values come directly from the components of the generalist GMM capturing the experiment's $\alpha$- and $K$-distributions (Section \ref{sec:EnsemblePreds}). 
When there were more predicted GMM components than experimental states in the model being considered, the trajectory ensemble generated for training used random subsets of all the predicted components.
For example, in a DI experiment with exactly two states and a GMM with five predicted components, each FOV generated for training would use a random set of two components from the five predicted.

\section{Results}\label{sec:Results}

Inspired by the 2024 AnDi Challenge, we evaluate U-AnD-ME using a well-balanced dataset with approximately 216,000 trajectories representative of a wide range of biologically relevant phenomena (Section \ref{sec:Methods_AnDiData}, Table \ref{tab:Exps}).
We first evaluate how impactful the framework's experiment-specific training is (Section \ref{sec:gen_vs_spec}).
We then discuss the quality of the single-trajectory analysis, including the change-point detection, the inference of the diffusion properties (the anomalous-diffusion exponent $\alpha$ and the generalized diffusion coefficient $K$, Section \ref{sec:Diff_Props}), and the classification of the diffusion type (Section \ref{sec:DTClassification}). Finally, we discuss the ensemble predictions (Section \ref{sec:EnsProps}).

\begin{figure*}[ht]
\centering
\includegraphics[width=0.95\linewidth]{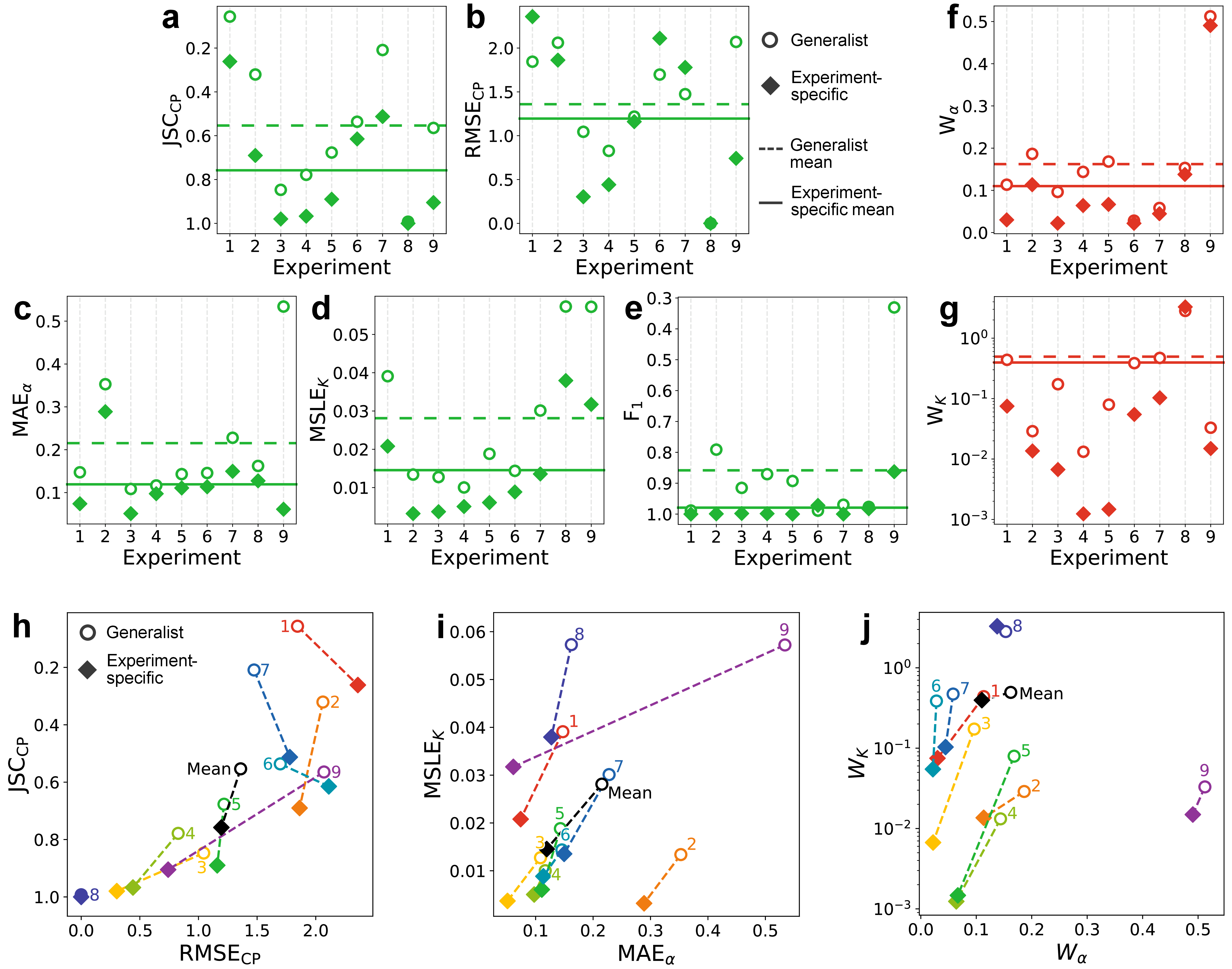}
\caption{\textbf{Impact of experiment specificity.}
(\textbf{a-e}) Performance by experiment (numbers as in Section \ref{sec:Methods_AnDiData}) for each metric of the single-trajectory analysis: (\textbf{a}) ${\rm JSC}_{\rm CP}$, (\textbf{b}) ${\rm RMSE}_{\rm CP}$, (\textbf{c}) ${\rm MAE}_{\alpha}$, (\textbf{d}) ${\rm MSLE}_{K}$, and (\textbf{e}) ${\rm F}_{1}$ score. (\textbf{f-g}) Performance by experiment for each metric of the ensemble analysis: (\textbf{f}) ${W}_{\alpha}$, and (\textbf{j}) ${W}_{K}$.
The $y$-axes show better metric values at the bottom of each plot.
Dashed and solid horizontal lines show average performance for generalist and experiment-specific networks, respectively.
(\textbf{h-j}) Correlation between subtask metrics for (\textbf{h}) change-point detection, (\textbf{i}) inference of diffusion properties and (\textbf{j}) ensemble predictions. The numbers indicate individual experiments. Dashed lines connect each experiment's generalist and experiment-specific metrics' values. Mean values are shown in black. Axes show better metric values at the bottom left of each plot.
(\textbf{a-j}) Hollow circles and filled diamonds represent metrics values of generalist and experiment-specific networks, respectively.
}
\label{fig:Exp_Spec}
\end{figure*}

\subsection{Generalist vs. Experiment-Specific Networks}\label{sec:gen_vs_spec}
Fig. \ref{fig:Exp_Spec} compares the performance of the generalist network to that of experiment-specific ones which leverage ensemble predictions.
For all metrics, experiment-specific networks led to improvements in average performance (Fig. \ref{fig:Exp_Spec}a-g), highlighting their benefit over the generalist because of a more relevant selection of diffusion properties for training. In most cases, there is also a strong correlation between the improvement of paired metrics going from generalist to experiment-specific networks (Fig. \ref{fig:Exp_Spec}h-j): for change-point detection, an improved ${\rm JSC}_{\rm CP}$ typically comes with a lower ${\rm RMSE}_{\rm CP}$ (Fig. \ref{fig:Exp_Spec}h); a similar trend emerges for the joint predictions of the trajectories' diffusion properties (Fig. \ref{fig:Exp_Spec}i) and of their ensemble distributions (Fig. \ref{fig:Exp_Spec}j). 
Only the ${\rm RMSE}_{\rm CP}$ for Experiments 1, 6 and 7, $W_K$ for Experiment 8 (due to its broad range of $K$ values)
and the ${\rm F}_1$-score for Experiment 6 deviate from this trend, with the decrease seen for this last metric being negligible.

For ${\rm RMSE}_{\rm CP}$, experiment specificity led to worse performance for Experiments 1, 6 and 7, as they contain more complex change points to detect due to multi-state diffusion or dimerization. As ${\rm RMSE}_{\rm CP}$ is calculated only on true positive change points (TP), the better ${\rm RMSE}_{\rm CP}$ less specific networks exhibit is not caused by superior change-point detection but by them simply failing to detect more difficult change points entirely.
Generalist networks detect only more distinct change points and make good predictions on these with relatively low localization error (hence the relatively higher values of ${\rm RMSE}_{\rm CP}$), while experiment-specific networks additionally pick up more subtle change points that are harder to localize well (hence the relatively lower values of ${\rm RMSE}_{\rm CP}$). This can lead to an anti-correlation between ${\rm JSC}_{\rm CP}$ and ${\rm RMSE}_{\rm CP}$ where favourable (high) ${\rm JSC}_{\rm CP}$ can be accompanied by poor (high) ${\rm RMSE}_{\rm CP}$, as in Fig \ref{fig:Exp_Spec}h for Experiments 1, 6 and 7.
This counter-intuitive behaviour is why both ${\rm RMSE}_{\rm CP}$ and ${\rm JSC}_{\rm CP}$ were used in tandem to evaluate change-point detections in the 2024 AnDi Challenge \cite{AnDi_2024}.

Interestingly, for Experiment 8 (serving as negative control to the other experiments \cite{AnDi_2024}), the values of most metrics changed little between generalist and experiment-specific networks. In particular, the latter led only to negligible improvements for ${\rm JSC}_{\rm CP}$, ${\rm RMSE}_{\rm CP}$, and ${\rm F}_{1}$-score. The relatively simpler dynamics of the SS model (single diffusion state with no change points) meant that even our generalist network could get close to optimal values for these metrics, thus reducing the need for experiment-specificity. 
Experiment 9 instead stands out due to the improvement that experiment-specific networks introduced over the generalist one on all metrics of the single-trajectory task (Fig. \ref{fig:Exp_Spec}a-e,h-i). This experiment in fact possesses two diffusion states that have very different properties (trapped and near ballistic)  with narrow distributions (Table \ref{tab:Exps}), thus justifying the greatest benefits over other experiments introduced by a better selection of parameters due to experiment-specific training.

\subsection{Change-Point Detection}\label{sec:change points}

\begin{figure*}[ht]
\centering
\includegraphics[width=\linewidth]{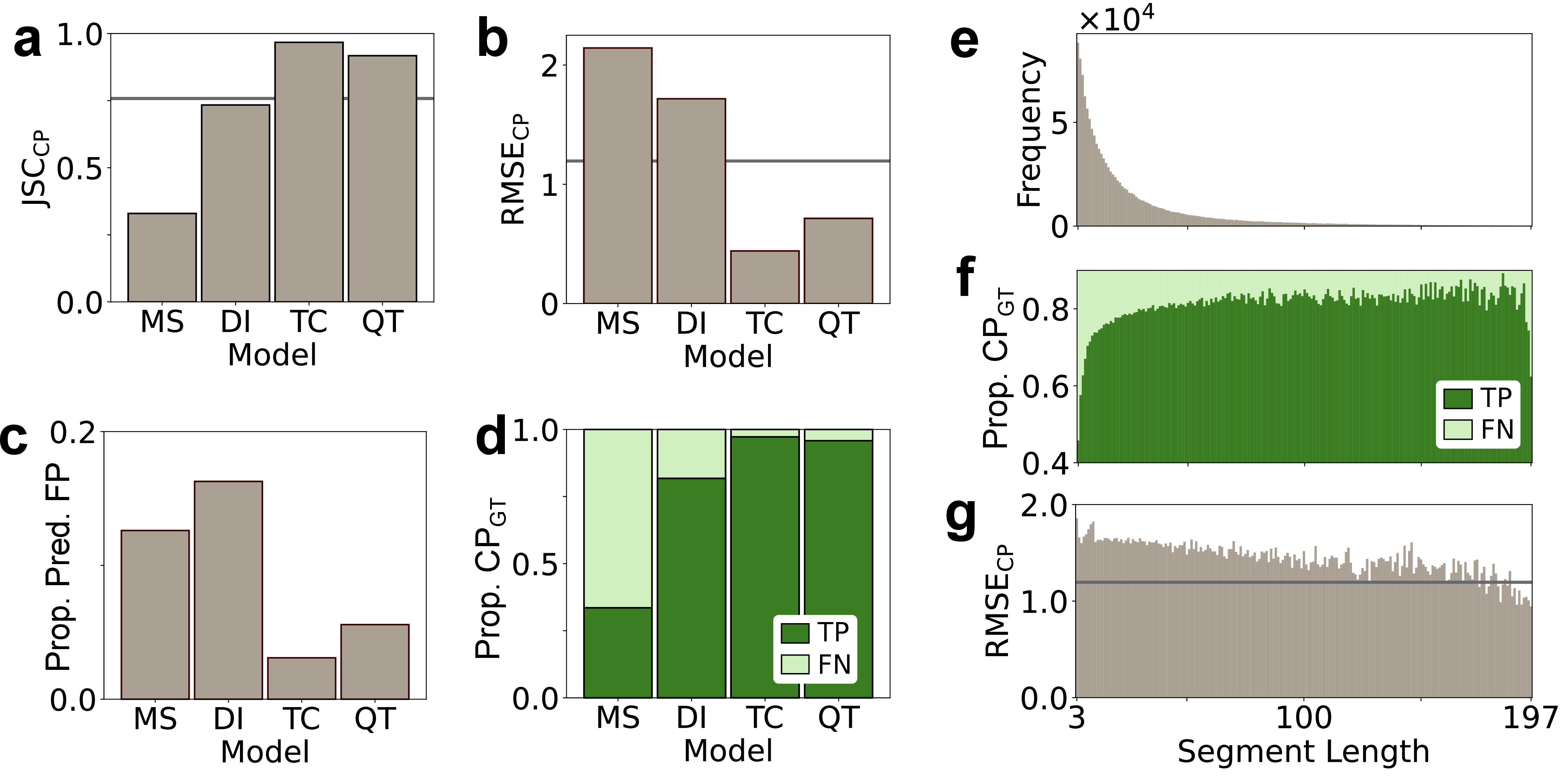}
\caption{\textbf{Change-point detection}.
(\textbf{a}-\textbf{b}) Metrics for change-point detection from experiment-specific networks for the four models that exhibit change points (MS, DI, TC and QT) in terms of (\textbf{a}) JSC\textsubscript{CP} and (\textbf{b}) RMSE\textsubscript{CP}. Horizontal lines represent the average value of each metric. (\textbf{c}) The proportion of false positives (FP) over all detected change points. 
(\textbf{d}) The relative proportion of true positives (TP, dark green) and false negatives (FN, light green) over all ground-truth change points (CP\textsubscript{GT}). 
(\textbf{e}) The frequency of different segment lengths across our dataset. Shorter segments are more represented as predictions are more challenging due to the lower information content per segment.
(\textbf{f}) The relative proportion of true positives (TP, dark green) and false negatives (FN, light green) across all ground-truth change points (${\rm CP}_{\rm GT}$ as in \textbf{d}) by segment length.
(\textbf{g}) ${\rm RMSE}_{\rm CP}$ of the change-point detections by segment length.   
Every change point is assigned to two segments, the ones immediately preceding and following it. The horizontal line shows the metric's average value. 
}
\label{fig:ChangePoints}
\end{figure*}

U-AnD-ME outperformed all other methods in the 2024 AnDi Challenge in terms of change-point detection, scoring better in terms of both $\rm JSC_{\rm CP}$ (accuracy) and $\rm RMSE_{\rm CP}$ (localization) \cite{AnDi_2024}.
The JSC\textsubscript{CP} (Fig. \ref{fig:ChangePoints}a) and the corresponding RMSE\textsubscript{CP} (Fig. \ref{fig:ChangePoints}b) show that change-point detection was highly performant, being both accurate (JSC\textsubscript{CP}) and precise (RMSE\textsubscript{CP}), for two models (TC and QT) with ${\rm JSC}_{\rm CP} > 0.91$ and a ${\rm RSME}_{\rm CP} < 0.72$ time steps. As a reference, all submitted methods in the segmentation task of the previous 2020 AnDi Challenge obtained ${\rm RSME}_{\rm CP}$ values of at best 10-20 time steps (an order of magnitude larger) when only considering the subset of trajectories with change points at least 20 time steps away from their start/end \cite{andiPython}. 
U-AnD-ME's particularly high performance for TC and QT is due to these two models showing two very clearly different states.
Unlike the other models with change points, TC and QT transitions always involve one segment with very low mobility -- near zero for TC and precisely zero for QT. This leads to more distinct change points, reducing mislabelling from the network, as can be seen from the relatively low number of false-positive (FP, Fig. \ref{fig:ChangePoints}c) and false-negative (FN, Fig. \ref{fig:ChangePoints}d)  detected change points compared to the other two models (MS and DI). This common segment feature in the change points may also facilitate a more effective training generalization among different trajectories. 

The MS model shows the poorest performance with ${\rm JSC}_{\rm CP} > 0.32$ and a ${\rm RSME}_{\rm CP} < 2.15$ time steps as, unlike other models, the change points in MS trajectories can involve a variety of states, which naturally adds complexity to their identification. Also, unlike TC or QT, these transitions can be between high mobility states. The relatively poorer JSC\textsubscript{CP} for MS is caused both by change points being missed (low TP and high FN, Fig. \ref{fig:ChangePoints}d) and by a relatively high proportion of false change points being identified (FP, \ref{fig:ChangePoints}c).

DI, with only two diffusion states as TC and QT, fares in-between as the change in diffusion properties is not always as marked as for these other two models (Table \ref{tab:Exps}). Like MS, DI can also include transitions between high-mobility states, but, differently from this model, penalization in detection comes from a higher proportions of false positive values rather than disproportionally mislabelling true change points (Fig. \ref{fig:ChangePoints}c-d). 

Figs. \ref{fig:ChangePoints}e-g explore the influence of segment length on change point detection.
As in the 2024 Challenge, the evaluation dataset is richer in shorter segments (Fig. \ref{fig:ChangePoints}e); this mirrors the fact that single-molecule live-imaging data often contain short trajectories. 
Additionally, this segment length distribution allows training to focus on identifying and characterising shorter segments, which are known to be more challenging and easier to miss due to their lower information content \cite{AnDi_2020,UAnDi}.
As can be expected, prediction quality increases with segment length (Fig. \ref{fig:ChangePoints}f-g), confirming that the improved feature information provided by longer trajectories has a notable impact on change-point identification \cite{AnDi_2020}. U-AnD-ME struggles most for very short segments ($< 10$ time steps, Fig. \ref{fig:ChangePoints}f-g). The proportion of true positive (TP) change points over false negatives (FN) across all ground-truth values increases steeply for increasing lengths, until it plateaus at approximately 83\% for segments longer than 65 time steps (Fig. \ref{fig:ChangePoints}f). Similarly, the ${\rm RMSE}_{\rm CP}$ shows a roughly linear halving from 1.86 to 0.94 time steps with increasing segment lengths (Fig. \ref{fig:ChangePoints}g).  
Interestingly, for very long segments ($> 194$ time steps), while the localization error is low (${\rm RMSE}_{\rm CP} < 1$ time steps (Fig. \ref{fig:ChangePoints}g), there is a higher proportion of false negative points compared to slightly shorter segments (Fig. \ref{fig:ChangePoints}g). 
This is an artefact coming from the high proportion of false negative points identified on very short segments. 
As the maximum trajectory length is 200, any detection shortcoming in trajectories with a single change point associated with a very short segment is also bound to propagate to its longer counterpart \cite{AnDi_2020}. 

\subsection{Inference of the Diffusion Parameters}\label{sec:Diff_Props}

\begin{figure*}[ht]
\centering
\includegraphics[width=0.8\linewidth]{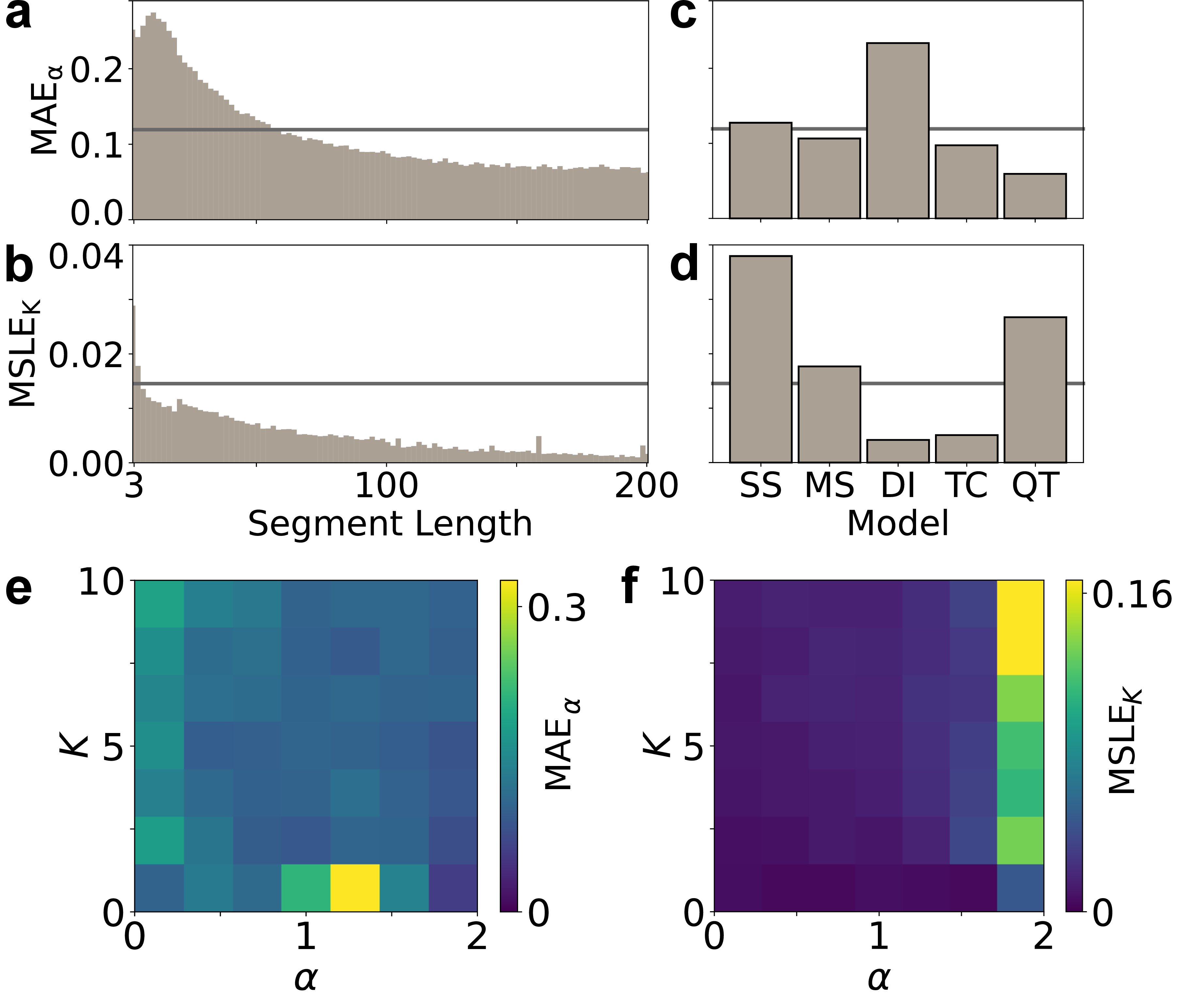}
\caption{\textbf{Inference of the diffusion properties.} 
(\textbf{a}) MAE\textsubscript{$\alpha$} and (\textbf{b}) MSLE\textsubscript{$K$} as a function of ground-truth segment length. 
(\textbf{c}) MAE\textsubscript{$\alpha$} and (\textbf{d}) MSLE\textsubscript{$K$} for each of the five possible diffusion models. Horizontal lines show average values. 
(\textbf{e-f}) Heat-maps showing (\textbf{e}) MAE\textsubscript{$\alpha$} and (\textbf{f}) MSLE\textsubscript{$K$} as a function of $\alpha$ and $K$. 
}
\label{fig:Diff_Props}
\end{figure*}

U-AnD-ME also outperformed all other participating methods in the 2024 AnDi Challenge in terms of inferring both $\alpha$ and $K$, achieving the lowest MAE\textsubscript{$\alpha$} and MSLE\textsubscript{$K$} \cite{AnDi_2024}. 
On our dataset, we achieved averages of $\rm{MAE}_\alpha=0.12$ and ${\rm MSLE}_K=0.015$. Inference strongly improves with segment length, due to the aforementioned increased feature information that longer segments contain (Figs. \ref{fig:Diff_Props}a-b) \cite{AnDi_2020}. After an initial steep improvement with segment length, $\rm{MAE}_{\alpha}$ plateaus at $\approx 0.7$ for segments longer than 150 time steps, while ${\rm MSLE}_{K}$ keeps slowly improving (Figs. \ref{fig:Diff_Props}b).

When evaluating the inference of $\alpha$ by model (Figs. \ref{fig:Diff_Props}c), ${\rm MAE}_{\rm \alpha}$ for SS, MS, and TC are all comparable and close to the average value. 
QT shows the best performance (${\rm MAE}_{\rm \alpha} = 0.06$), likely because inferring $\alpha$ is straightforward for the immobilized state ($\alpha=0$). The inference of segment properties can also be expected to be most effective when change points are accurately detected (Fig. \ref{fig:ChangePoints}), which slightly benefited TC too (Section \ref{sec:change points}).  
At the opposite end, DI was the worst performing model (${\rm MAE}_{\rm \alpha} = 0.23$).
This is attributed largely to Experiment 2 which, unlike the two other DI experiments (Experiments 5 and 6) has two overlapping states (diffusion and directed) with very similar values of $\alpha$, leading to mislabelling the less frequent directed state (Table \ref{tab:Exps}).
In fact, this experiment has the worst ${\rm MAE}_\alpha$ of all even after using experiment-specific networks (Fig. \ref{fig:Exp_Spec}c). Interestingly, the inference of $K$ by model (Figs. \ref{fig:Diff_Props}d) shows much more variability than that of $\alpha$.
MS performance is the only one close to the MSLE\textsubscript{K} average value, with the other models performing much better or worse than average. TC and DI performed best, both with MSLE\textsubscript{K} scores of ca. 0.005, as all experiments of these models have well-separated bimodal distributions of $K$ (Table \ref{tab:Exps}, Section \ref{sec:EnsProps}), which U-AnD-ME could discern well. On the opposite end, SS performed the worst, likely due to the range of $K$ values in its only experiment (the negative-control Experiment 8) being significantly broader than any other experiment. 
Although better than SS, QT performance was also poorer than the average, hampered by Experiment 9 with an extremely superdiffusive state ($\mu_\alpha=1.99$). We believe that its very directed trajectories have indeed influenced our network's capability to finely resolve the correct width of the distribution of $K$-values (see also Section \ref{sec:EnsProps}).

Finally, Figs. \ref{fig:Diff_Props}e-f show how the values of MAE\textsubscript{$\alpha$} and MSLE\textsubscript{$K$} vary with $\alpha$ and $K$. Performance is generally quite consistent across different values with two exceptions: MAE\textsubscript{$\alpha$} at $\alpha \approx 1.2$ for low values of $K$ ($K < 1.5$) (Fig. \ref{fig:Diff_Props}e) and ${\rm MSLE}_K$ for strongly superdiffusive trajectories as $K$ grows (Fig. \ref{fig:Diff_Props}f). Both deviations are however not due to the actual values of $\alpha$ and $K$, but rather to their representation in the training dataset and the shape of the experimental distributions to be resolved by U-AnD-ME (Section \ref{sec:EnsProps}). In fact, while the latter is due to Experiment 8 (the negative-control experiment) and its atypically broad range of $K$-values compared to the other experiments (Table \ref{tab:Exps}), the former is due to U-AnD-ME struggling to resolve the multimodal distributions of Experiments 6 and 7, both featuring a secondary narrow peak at $\alpha=1.2$ on a much broader underlying distribution centered at $\alpha=1$ (Table \ref{tab:Exps}, Section \ref{sec:EnsProps}).  \\

\subsection{Classification of Diffusion Type}\label{sec:DTClassification}

\begin{figure*}[ht]
\centering
\includegraphics[width=\linewidth]{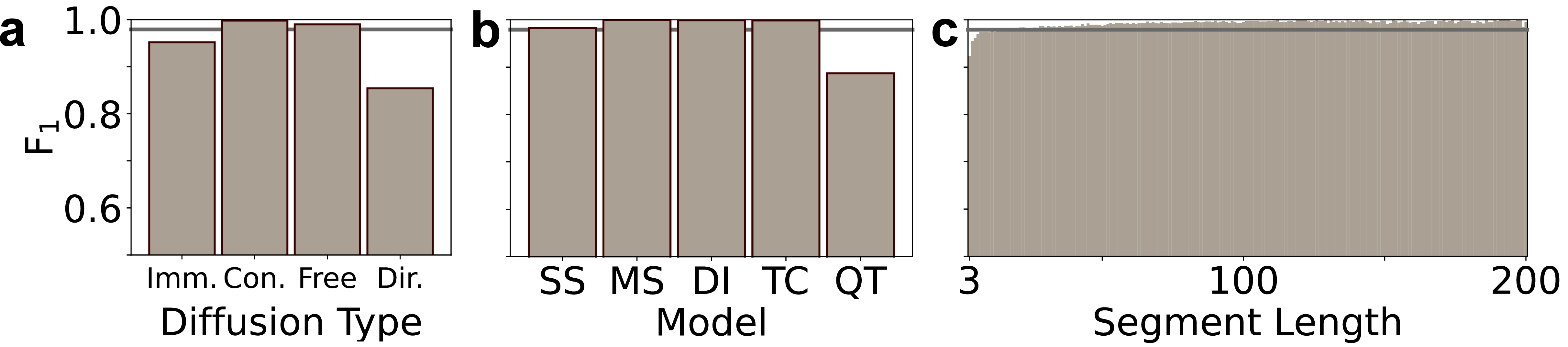}
\caption{\textbf{Classification of Diffusion Type.}
F\textsubscript{1}-score of the classification of diffusion type (DT) by (\textbf{a}) diffusion type (immobilized, confined, freely diffusing, or directed), (\textbf{b}) diffusion model, and (\textbf{c}) ground-truth segment length. The solid lines show the average F1 score across all experiments.}
\label{fig:state}
\end{figure*}

When it came to the classification of diffusion type (among immobilized, confined, freely diffusing and directed), U-AnD-ME achieved a superior F\textsubscript{1}-score (with an average score of $\rm{F}_{1}=0.98$) compared to all other participating methods in the 2024 AnDi Challenge \cite{AnDi_2024}. Fig. \ref{fig:state}a shows that U-AnD-ME was indeed highly accurate in classifying all diffusion types, being the lowest F\textsubscript{1} score 0.85 for directed diffusion. Looking at the classification by model (Fig. \ref{fig:state}b) shows how this relatively poorer performance for directed trajectories is primarily due to the QT model. As noted earlier (Section \ref{sec:Diff_Props}), this model was hampered by Experiment 9 with an extremely superdiffusive state that was harder to resolve for U-AnD-ME (see also Fig. \ref{fig:Exp_Spec}e). Finally, classification as a function of segment length is unsurprising, with relatively poorer predictions for shorter segments due to their lower information content (Fig. \ref{fig:state}c). Segments of length 3 achieved F\textsubscript{1}=0.92, with values plateauing to ${\rm F}_1 \approx 1$ for segments longer than 100 time steps.

\subsection{Ensemble Distributions}\label{sec:EnsProps}

\begin{figure*}[ht]
\centering
\includegraphics[width=0.8\linewidth]{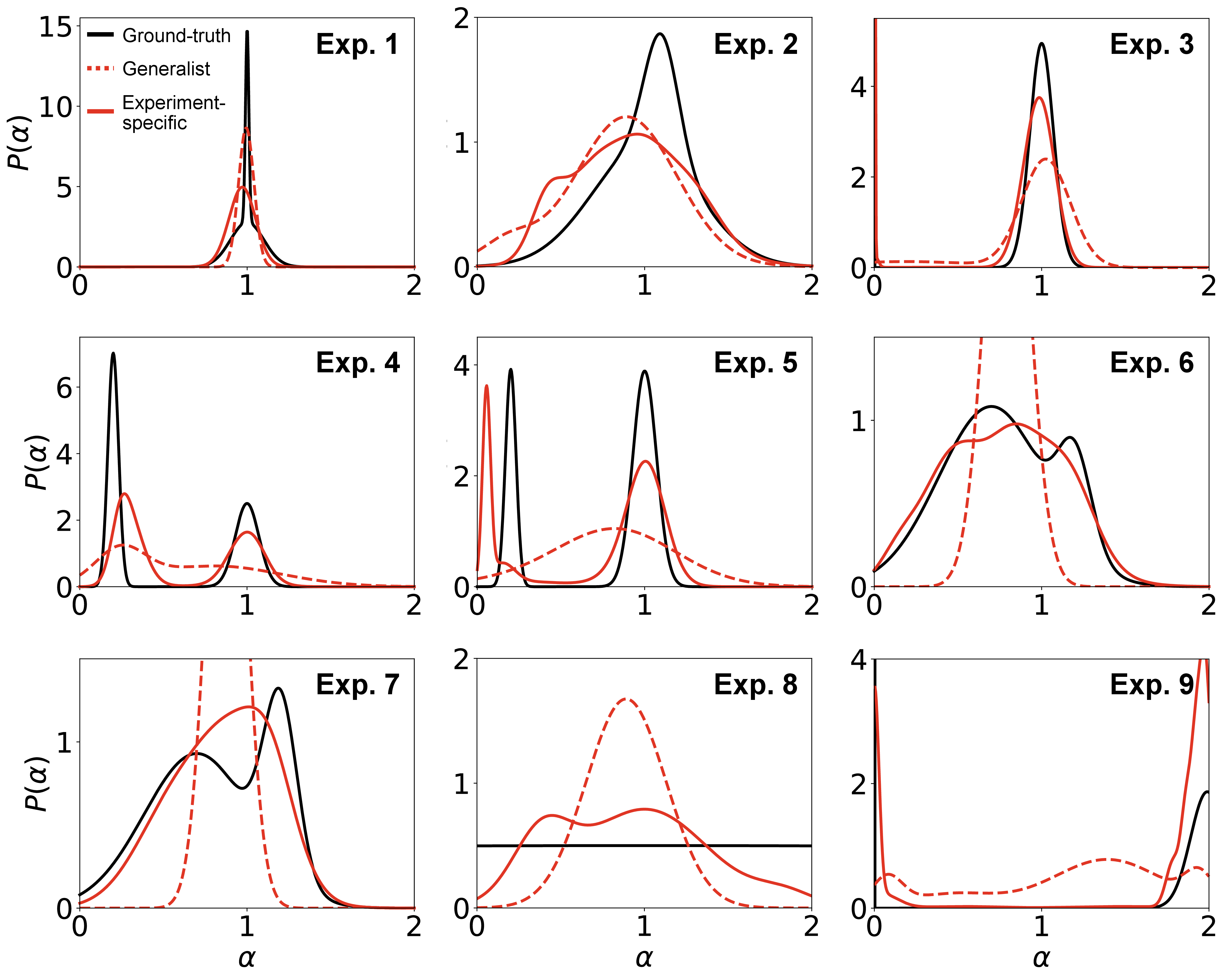}
\caption{\textbf{Predicted $\alpha$-distributions from ensemble analysis.} Ground-truth $\alpha$-distributions $P(\alpha)$ (black lines), distributions predicted by generalist networks (dashed red lines), and distributions predicted by experiment-specific networks (solid red lines) for each experiment.}
\label{fig:ens_alpha}
\end{figure*}

\begin{figure*}[ht]
\centering
\includegraphics[width=0.8\linewidth]{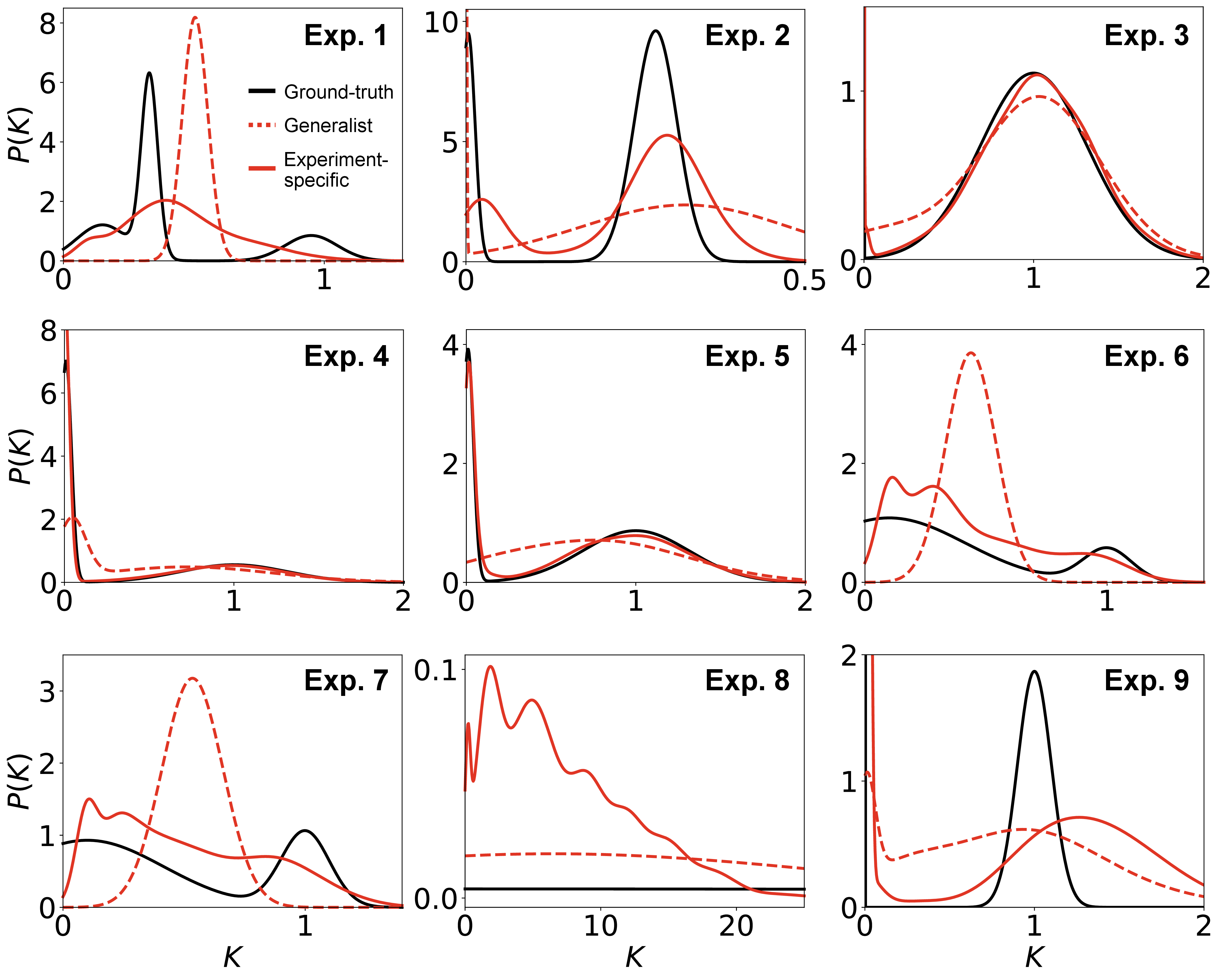}
\caption{\textbf{Predicted $K$-distributions from ensemble analysis.} Ground-truth $K$-distributions $P(K)$ (black lines), distributions predicted by generalist networks (dashed red lines), and distributions predicted by experiment-specific networks (solid red lines) for each experiment.}
\label{fig:ens_K}
\end{figure*}

Finally, U-AnD-ME outperformed all other participating methods in the 2024 AnDi Challenge in terms of capturing the ensemble distribution of $\alpha$. $W_\alpha$ values in Fig. \ref{fig:Exp_Spec}f show an average of 0.16 and 0.11 for generalist and experiment-specific networks, respectively.
Our method was not as successful in terms of predicting the distribution of $K$, placing 4\textsuperscript{th}; this was the only metric across the Trajectory Track of the Challenge for which U-AnD-ME did not outperform all other participating methods.
We primarily attribute this decreased performance to the negative-control Experiment 8 (Section \ref{sec:Diff_Props}), which shows the worst $W_K$ in Fig. \ref{fig:Exp_Spec}g with scores of 2.83 and 3.29 against averages of 0.49 and 0.40 for  generalist and experiment-specific networks, respectively.

From the reconstructed ensemble probabilities of $\alpha$ and $K$ in Figs. \ref{fig:ens_alpha} and \ref{fig:ens_K}, we can see how experiment-specific networks tend to perform better than generalist ones as already observed in Fig. \ref{fig:Exp_Spec}f-g,j. For both properties, generalist networks tend to find unimodal distributions near the average parameter value, while experiment-specific networks capture richer distributions better reflecting those of the ground truth. When two peaks in the ground-truth distributions are close together (e.g., in the $\alpha$-distributions for Experiments 2, 6 and 7, and in the $K$-distribution for Experiment 1), experiment-specific networks tend to approximate neighbouring ground-truth modes with a single broad coarse-grained peak, merging the information from the individual modes. Unimodal distributions (e.g., in the $\alpha$-distributions for Experiments 1 and 3, and in the $K$-distribution for Experiments 3) and multimodal distributions formed by well-separated modes (e.g., in the distributions of $\alpha$ for Experiments 4, 5 and 9, and of $K$ for Experiments 2, 4, 5, 6, 7 and 8) tend instead to be captured better by U-AnD-ME.

\section{Discussion}\label{sec:Discussion}

The success of U-AnD-ME in the 2024 AnDi Challenge demonstrates its significance and potential for the analysis of data from live-cell single-molecule imaging at both single-trajectory and ensemble levels.
Our method came 1\textsuperscript{st} for the Single-Trajectory Task of the Trajectory Track of the Challenge, which was the most subscribed task of the competition. For this highly competitive task, U-AnD-ME was awarded 1\textsuperscript{st} place for every possible subtask, accurately predicting the locations of change points, the anomalous-diffusion exponent $\alpha$, the generalized diffusion coefficient $K$, and the diffusion type. Our method was also awarded 1\textsuperscript{st} place for the Ensemble Task of the Trajectory Track, coming 1\textsuperscript{st} in the subtask dedicated to predicting the distribution of $\alpha$ and 4\textsuperscript{th} in the subtask for predicting the distribution of $K$. The suboptimal performance for this subtask is due to our framework struggling in cases where the training distribution for the generalist network differs significantly from the experimental distribution. In the Challenge, the generalised diffusion coefficient $K$ could take values in the range $K\in[10^{-12}, 10^6]$. As most experiments but the negative-control Experiment 8 generally take smaller values, the training procedure we used for generalist networks focussed on a smaller range ($K\in[10^{-12}, 15]$), as training a network over the full range would be computationally expensive and result in poor performance for the range of interest. As a consequence, experiments with significant density for $K>15$ may have poor generalist network predictions for the ensemble $K$-distribution, and thereby poor experiment-specific network predictions for this distribution. No such issue affects $\alpha$ as these values are confined to a much smaller interval ($\alpha\in[0,2)$), making training over all possible values straightforward. 
In practice, prior knowledge of the system being analysed could eliminate this issue by adapting the generalist network's training range to better suit the problem at hand.

The inference of segment diffusion properties tends to improve with the quality of the change-point detections \cite{AnDi_2020}.
Currently, U-AnD-ME uses a standard binary cross-entropy loss for change points. 
However, in typical trajectories, fewer time steps are change points than not, leading to significant imbalances between these two possible classes.
\textit{Focal loss} is a modified cross-entropy designed to perform better with class imbalance \cite{FocalLoss}. Its use for training instead of binary cross-entropy could improve the performance of change-point detection and thereby improve the inference of segment properties too.
Moreover, as shown here, experiment-specific training significantly improves the accuracy of all network predictions. Using experiment-specific architectures informed by the physics of each underlying model could further improve our approach. 
Ensemble predictions could also benefit from increased a priori knowledge about experimental data: knowing an experiment's model could inform us about the number of distinct states it has and this could be used to directly enforce the number of GMM components. Finally, while the original U-Net 3+ architecture was designed for image analysis \cite{UNet3+}, U-AnD-ME extended it to time series.
Amongst other changes, this involved using 1D convolutions as opposed to 2D convolutions. 
Presently, U-AnD-ME analyses trajectories extracted from microscopy videos. 
Future developments could explore the use of 3D convolutions to enable U-AnD-ME to directly extract information from these videos.

In conclusion, built using convolutional operations, our machine learning framework ensures high efficiency and stable training while delivering results that allowed U-AnD-ME to be among the top-performing teams in the 2024 Andi Challenge. Even against comparable frameworks for the analysis of anomalous diffusion data proposed during the 2020 AnDi Challenge \cite{AnDi_2020}, U-AnD-ME offers several advantages as it does not require complex feature engineering \cite{CONDOR}, uses a single network for a wide range of segment lengths \cite{RANDI, WADNet}, and extracts all diffusion parameters with a single architecture \cite{RANDI}.
The relative simplicity of our framework is particularly noteworthy given the increased complexity of the 2024 AnDi Challenge tasks compared to those of the previous version.
While we applied it to the analysis of 2D trajectories, our architecture could be easily adapted to handle even higher dimensional trajectories.
Besides its proven effectiveness in particle-tracking analysis, U-AnD-ME’s central architecture could also be relevant for any other task requiring analysis of time series exhibiting complex diffusion behaviour, such as animal migration records \cite{Birds} or financial market data \cite{Econ_AnomDiff}. 
Additionally, it holds potential for any problem that requires the segmentation of time series.
For example, U-AnD-ME framework could be applied to ECG analysis \cite{ECG}, fault detection in manufacturing \cite{ManufacturingFault}, or detecting ecosystem regime shifts \cite{EcosystemSegment}.

\section*{Acknowledgements}
S.A. and G.V. are grateful for the studentship funded by the A*STAR-UCL Research Attachment Programme through the EPSRC M3S CDT (EP/L015862/1). 
R.N. acknowledges support from the Academic Research Fund from the Singapore Ministry of Education (RG59/21 and MOE2019-T2-2-010) and the National Research Foundation, Singapore, under its 29th Competitive Research Program (CRP) Call (Grant No. NRF-CRP29-2022-0002). This work has also been supported by The Chan Zuckerberg Initiative “Multi-color single molecule tracking with lifetime imaging” (2023-321188). 

\section{Bibliography}
\bibliographystyle{ieeetr}

\end{document}